\documentclass[reprint,showpacs,preprintnumbers,superscriptaddress,nofootinbib,amsmath,amssymb,aps,prl,floatfix]{revtex4-1}
\pdfoutput=1
\usepackage{graphicx}
\usepackage{dcolumn}
\usepackage{bm}
\usepackage{url}
\usepackage[ddmmyy,24hr]{datetime}
\usepackage{varwidth}
\usepackage{multirow}
\newcommand{\ns}{\Delta{N}_{\text{eff}}}
\begin{document}

\preprint{\begin{tabular}{l}
\texttt{arXiv:1404.6160 [astro-ph.CO]}
\end{tabular}}

\title{Cosmological Invisible Decay of Light Sterile Neutrinos}

\author{S. Gariazzo}
\affiliation{Department of Physics, University of Torino, Via P. Giuria 1, I--10125 Torino, Italy}
\affiliation{INFN, Sezione di Torino, Via P. Giuria 1, I--10125 Torino, Italy}

\author{C. Giunti}
\affiliation{INFN, Sezione di Torino, Via P. Giuria 1, I--10125 Torino, Italy}

\author{M. Laveder}
\affiliation{Dipartimento di Fisica e Astronomia ``G. Galilei'', Universit\`a di Padova,
and
INFN, Sezione di Padova,
Via F. Marzolo 8, I--35131 Padova, Italy}


\begin{abstract}
We introduce a cosmological invisible decay of the sterile neutrino with the eV-scale mass
indicated by short-baseline neutrino oscillation experiments
in order to allow its full thermalization in the early Universe.
We show that the fit of the cosmological data is practically as good as
the fit obtained with a stable sterile neutrino without mass constraints,
which has been recently considered by several authors
for the explanation of the observed suppression of small-scale matter density fluctuations
and
for a solution of the tension between the Planck and BICEP2 measurements of
the tensor-to-scalar ratio of large-scale fluctuations.
Moreover,
the extra relativistic degree of freedom corresponding to a fully thermalized sterile neutrino
is correlated with a larger value of the Hubble constant, which is in agreement with
local measurements.
\end{abstract}

\pacs{14.60.Pq, 14.60.Lm, 14.60.St, 98.80.-k}

\maketitle

The recent results of the BICEP2 experiment \cite{Ade:2014xna}
revived the interest in the cosmological contribution of light sterile neutrinos
\cite{Giusarma:2014zza,Zhang:2014dxk,Dvorkin:2014lea,Archidiacono:2014apa,Zhang:2014nta}.
BICEP2 measured a tensor-to-scalar ratio
$r = 0.20 {}^{+0.07}_{-0.05}$
of large-scale fluctuations,
which is significantly larger than
the upper limits of
WMAP \cite{Hinshaw:2012aka} ($r < 0.13$ at 95\% CL)
and
Planck \cite{Ade:2013zuv} ($r < 0.11$ at 95\% CL).
This tension can be relieved by allowing an extra relativistic particle
in the early Universe,
which could be a sterile neutrino
\cite{Giusarma:2014zza,Zhang:2014dxk,Dvorkin:2014lea,Archidiacono:2014apa,Zhang:2014nta}.
The effect is due to the correlation between
the effective number of relativistic degrees of freedom
$N_{\text{eff}}$
before photon decoupling
(see \cite{Archidiacono:2013fha,Lesgourgues:2014zoa})
and the spectral index $n_s$
of the scalar primordial curvature power spectrum
$
\mathcal{P}_{\mathcal{R}}(k)
=
A_{s} (k/k_{0})^{n_{s}-1}
$,
with the pivot scale $k_{0} = 0.05 \, \text{Mpc}^{-1}$,
which lies roughly in the middle of the logarithmic range of scales probed by Planck
\cite{Ade:2013zuv}.
Keeping fixed the amplitude $A_{s}$ at $k \sim k_{0}$,
which is constrained by the high-precision Planck data,
the scalar contribution to the
large-scale temperature fluctuations with $k \ll k_{0}$
measured by WMAP and Planck
can be decreased by an increase\footnote{
One could think to
alleviate the tension between BICEP2 and WMAP-Planck
by decreasing $n_s$,
if the value of $r$ measured by BICEP2
refers to a wavenumber $k_{1}$
larger than than the wavenumber
$k_{2} = 0.002 \, \text{Mpc}^{-1}$
corresponding to the WMAP and Planck upper bounds
\cite{Ashoorioon:2014nta,Audren:2014cea}.
Since
$
r_{k_{2}}
\simeq
r_{k_{1}}
\left( k_{1} / k_{2} \right)^{n_{s}-1-n_{t}}
$,
where $n_{t}$ is the tensor spectral index,
for $k_{2} < k_{1}$
and
$n_{s}-1-n_{t} < 0$
we have $r_{k_{2}} < r_{k_{1}}$
and the ratio
$r_{k_{2}} / r_{k_{1}}$
decreases by decreasing $n_s$.
However,
one must take into account that WMAP and Planck
did not measure directly the tensor fluctuations as BICEP2,
but measured the temperature fluctuations,
in which the scalar and tensor contributions are indistinguishable.
Hence,
decreasing $n_{s}$ increases the scalar contribution to the
temperature fluctuations measured by
WMAP and Planck at $k_{2} < k_{1}$
and there is less room for a tensor contribution.
Therefore the WMAP and Planck upper bounds on $r_{k_{2}}$
tighten by about the same amount of the decrease of the BICEP2 value of
$r_{k_{2}}$,
maintaining the tension.
}
of the spectral index $n_s$.
In this way,
the WMAP and Planck data leave more space for the tensor contribution
\cite{Knox:1994qj}
and the corresponding bounds on $r$ are relaxed.
However,
the increase of $n_s$ induces an increase of small scale fluctuations
with $k \gg k_{0}$,
which would spoil the fit of high-$\ell$ Cosmic Microwave Background (CMB) data
if the increase is not
compensated by an effect beyond the standard cosmological $\Lambda$CDM model.
An increase of $N_{\text{eff}}$ above the Standard Model value
$N_{\text{eff}}^{\text{SM}} = 3.046$ \cite{Mangano:2005cc}
has just the desired effect of decreasing small scale fluctuations
(see Ref.~\cite{Hou:2011ec}).
For example,
from the fit of CMB data without BICEP2
the authors of Ref.~\cite{Archidiacono:2014apa}
obtained
$n_{s} = 0.970 {}^{+0.011}_{-0.018}$ ($1\sigma$)
and
$\Delta{N}_{\text{eff}} < 1.18$ ($2\sigma$),
with $\Delta{N}_{\text{eff}} = N_{\text{eff}} - N_{\text{eff}}^{\text{SM}}$,
and adding BICEP2 data they found
$n_{s} = 0.986 {}^{+0.016}_{-0.020}$ ($1\sigma$)
and
$\Delta{N}_{\text{eff}} = 0.82 {}^{+0.40}_{-0.57}$ ($1\sigma$).

A fit of cosmological data must take into account also
the measurements of the Hubble constant  $H_0$
and of the matter distribution in the Universe.
In particular,
the measured small-scale matter density fluctuations ($k \sim 0.1 \, \text{Mpc}^{-1}$)
are smaller than those obtained by evolving the primordial density fluctuations
with the relatively large matter density at recombination
measured precisely by Planck
(see Refs.~\cite{Wyman:2013lza,Battye:2013xqa,Gariazzo:2013gua}).
This discrepancy can be explained by the
existence of a massive sterile neutrino
with a free-streaming that suppresses the growth of structures
which are smaller than the free-streaming length
(see, for example, Ref.~\cite{Lesgourgues:2014zoa}).
A global fit of cosmological data gives
a sterile neutrino mass
$m_s = 0.44 {}^{+0.11}_{-0.16} \, \text{eV}$ ($1\sigma$),
with
$\Delta{N}_{\text{eff}} = 0.89 {}^{+0.34}_{-0.37}$ ($1\sigma$)
\cite{Archidiacono:2014apa}
(see also \cite{Zhang:2014dxk,Dvorkin:2014lea,Zhang:2014nta}).

The existence of light massive sterile neutrinos in the early Universe is
especially exciting in connection with the indications of short-baseline (SBL) neutrino oscillations
found in the LSND experiment
\cite{Aguilar:2001ty},
in Gallium experiments
\cite{Abdurashitov:2005tb}
and in reactor experiments
\cite{Mueller:2011nm,Mention:2011rk,Huber:2011wv}.
These oscillations can be explained
in the 3+1 extension of
the standard three-neutrino mixing paradigm
with the introduction of a sterile neutrino
with a mass
$m_s \sim 1 \, \text{eV}$
\cite{Kopp:2013vaa,Giunti:2013aea},
which is significantly larger than the cosmologically preferred value
mentioned above.
In fact,
such a large mass would induce an excessive suppression
of small-scale matter density fluctuations
if the sterile neutrinos are fully thermalized.

The authors of Ref.~\cite{Archidiacono:2014apa}
analyzed the cosmological data, including the BICEP2 results,
taking into account the global fit of neutrino oscillation data presented in Ref.~\cite{Giunti:2013aea}.
The result of the combined fit shows that a sterile neutrino with a mass at the eV scale
is allowed by cosmological data only if it is not fully thermalized in the early Universe:
$m_s = 1.19 {}^{+0.15}_{-0.12} \, \text{eV}$ ($1\sigma$)
and
$\Delta{N}_{\text{eff}} = 0.19 {}^{+0.07}_{-0.15}$ ($1\sigma$).
The case of a fully thermalized sterile neutrino is disfavored by $\Delta\chi^2>10$
\cite{Archidiacono:2014apa}.
Similar conclusions have been reached before
the recent BICEP2 results
(see the recent Refs.~\cite{Archidiacono:2013xxa,Mirizzi:2013kva,Gariazzo:2013gua},
which take into account the Planck data \cite{Ade:2013zuv})
and motivated the study of mechanisms which can suppress the
thermalization of sterile neutrinos in the early Universe
due to active-sterile oscillations before neutrino decoupling
\cite{Dolgov:2003sg,Cirelli:2004cz,Chu:2006ua,Hannestad:2012ky}.
Examples are a large lepton asymmetry
\cite{Hannestad:2012ky,Mirizzi:2012we,Saviano:2013ktj,Hannestad:2013pha},
an enhanced background potential due to new interactions in the sterile sector \cite{Hannestad:2013ana,Dasgupta:2013zpn,Bringmann:2013vra,Ko:2014bka,Archidiacono:2014nda},
a larger cosmic expansion rate at the time of sterile neutrino production
\cite{Rehagen:2014vna},
and
MeV dark matter annihilation
\cite{Ho:2012br}.

\begin{figure*}
\centering
\includegraphics[width=0.49\textwidth,page=1]{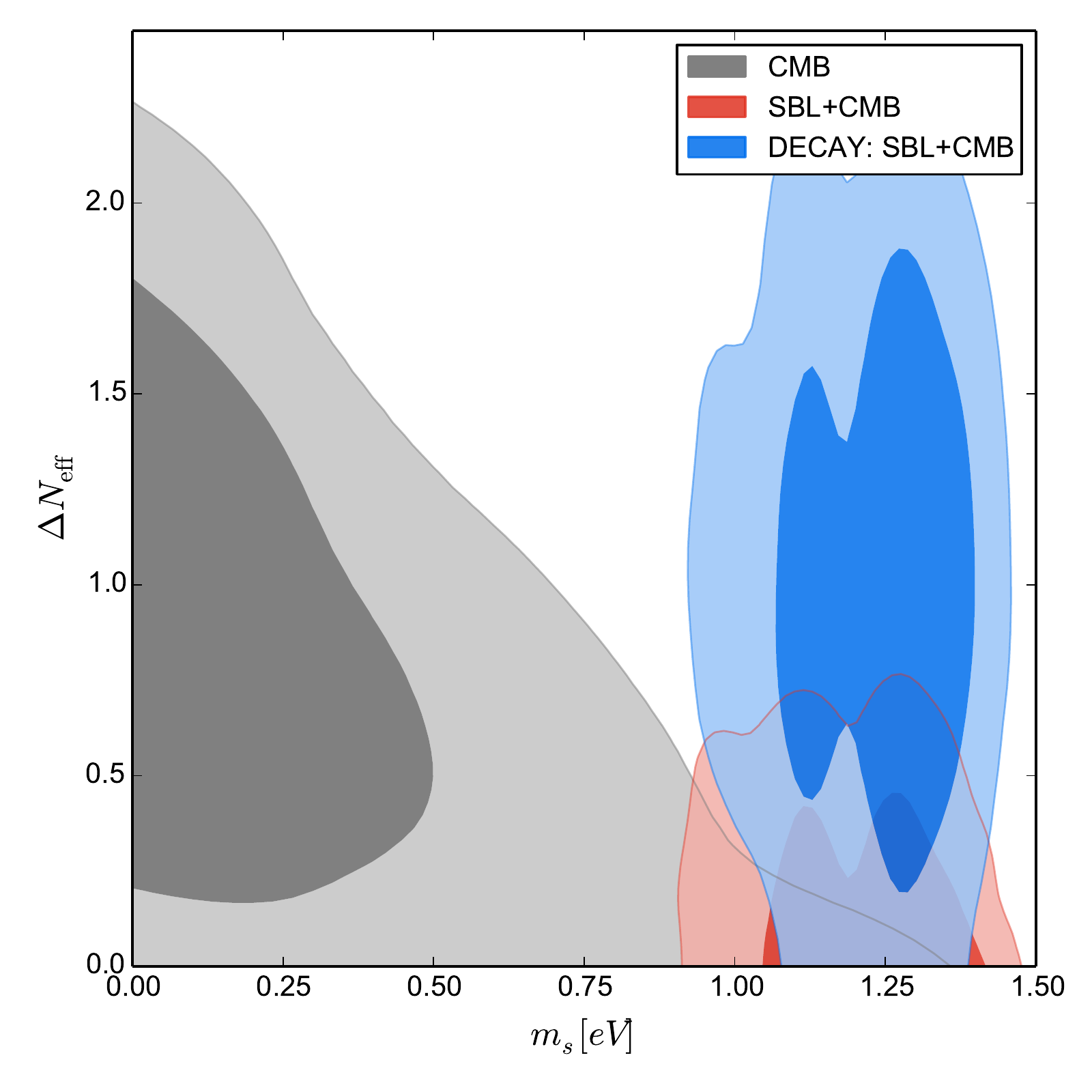}
\includegraphics[width=0.49\textwidth,page=2]{plots_decay.pdf}
\caption{\label{fig:cmb}
$1\sigma$ and $2\sigma$ marginalized allowed regions
obtained with CMB data (Planck+WP+high-$\ell$+BICEP2(9bins); see Ref.~\cite{Archidiacono:2014apa}).
The gray and red regions are those obtained in Ref.~\cite{Archidiacono:2014apa} without and with the SBL prior.
The blue regions are obtained by adding the invisible sterile neutrino decays.
}
\end{figure*}

In this letter we propose to solve the tension between the thermalization of the sterile neutrino in the early Universe
and the fit of cosmological data with a sterile neutrino having a eV-scale mass
by introducing an invisible decay of the sterile neutrino.
The decay must be invisible in order not to generate unobserved signals.
We assume that the decay products are very light or massless particles belonging to the sterile sector.
For example, the eV-scale sterile neutrino $\nu_{s}$ could decay into a lighter sterile neutrino $\nu_{s'}$
and a very light invisible (pseudo)scalar boson\footnote{
The new invisible light (pseudo)scalar boson
is assumed to interact only with the sterile neutrinos,
without the interactions with the active neutrinos
studied in Refs.~\cite{Beacom:2004yd,Hannestad:2005ex}
and references therein.
}
$\phi$.
The lighter sterile neutrino $\nu_{s'}$ must have very small mixing with the active neutrinos,
in order to forbid its thermalization in the early Universe and
to preserve the effectiveness of the standard three-neutrino mixing paradigm
for the explanation of solar and atmospheric neutrino oscillations.
Also the very light invisible boson $\phi$
has a negligible thermal distribution before the decay,
because it belongs to the sterile sector which
may have been in equilibrium at very early times,
but has decoupled from the thermal plasma at a very high temperature
and the densities of all the particles belonging to the sterile sector
have been washed out in the following phase transitions and heavy particle-antiparticle annihilations
(see, for example, Ref.~\cite{Dolgov:2002wy}).
Another possible decay which does not need the presence
of a light boson is
$\nu_{s} \to \nu_{s'} \bar\nu_{s'} \nu_{s'}$,
which needs an effective four-fermion interaction of sterile neutrinos.

In the invisible decay scenario,
the eV-scale sterile neutrino can be fully thermalized in the early Universe
through active-sterile oscillations
\cite{Dolgov:2003sg,Cirelli:2004cz,Chu:2006ua,Hannestad:2012ky}
and generates the $\Delta{N}_{\text{eff}} = 1$
indicated by the fit of CMB data with BICEP2.
In the first radiation-dominated part of the evolution of the Universe the mass of the sterile neutrino is not important,
because it is relativistic.
The mass effect is important in the following matter-dominated evolution of the Universe,
which leads to the formation of Large Scale Structures (LSS) and the current matter density.
The sterile neutrinos which decay into invisible relativistic particles
do not contribute to the matter budget.
In this way the eV-scale mass of the sterile neutrino
indicated by short-baseline oscillation experiments
becomes compatible with a full thermalization of the sterile neutrino in the early Universe.

We analyzed the same cosmological data considered in Ref.~\cite{Archidiacono:2014apa}
with a Bayesian analysis performed with the \texttt{CosmoMC} Monte Carlo Markov Chain engine
\cite{Lewis:2002ah}
modified in order to take into account the invisible decay of the sterile neutrino.
For simplicity\footnote{
A precise calculation requires the solution of the coupled Boltzmann equations
describing the evolution of the distributions of the sterile neutrino and the decay products.
The results of such a calculation in the framework of a specific decay model will be presented elsewhere.
},
we neglected the energy dependence of the sterile neutrino lifetime and we considered a sterile neutrino with a Fermi-Dirac distribution multiplied by
\begin{equation}
N_{s}(t)
=
\Delta{N}_{\text{eff}} \, e^{-t/\tau_{s}}
\,,
\label{Ns}
\end{equation}
where $t$ is the cosmic time and $\tau_{s}$ is the effective lifetime of the sterile neutrino.
For simplicity,
we neglect the energy distributions of the very light or massless invisible decay products
(which depend on the specific decay model)
and we parameterize their effect with an effective increase of the amount of radiation by
$\Delta{N}_{\text{eff}} \left( 1 - e^{-t/\tau_{s}} \right)$.
Following Refs.~\cite{Archidiacono:2013xxa,Gariazzo:2013gua,Archidiacono:2014apa},
we take into account the SBL constraint on $m_s$
through a prior given by the posterior of the global analysis of
SBL oscillation data presented in Ref.~\cite{Giunti:2013aea}.

Figure~\ref{fig:cmb}
shows the $1\sigma$ and $2\sigma$ marginalized allowed regions
in the planes
$m_s$--$\Delta{N}_{\text{eff}}$
and
$H_0$--$\Delta{N}_{\text{eff}}$
obtained by fitting the CMB data
(Planck+WP+high-$\ell$+BICEP2(9bins); see Ref.~\cite{Archidiacono:2014apa})
with the SBL prior
in a model with free $\Delta{N}_{\text{eff}}$
and a massive sterile neutrino which decays invisibly.
The corresponding numerical values of the cosmological parameters
are listed in Tab.~\ref{tab:all}.
From the values of
$\Delta\chi^2(\text{A})$ and $\Delta\chi^2(\text{B})$
one can see that the fit of cosmological data is even slightly better than that
obtained with a stable sterile neutrino without mass constraints (A)
and much better than that
obtained with a stable sterile neutrino with the SBL mass prior (B).

\begin{table}
\begin{center}
\renewcommand{\arraystretch}{1.4}
\begin{tabular}{|l|c|c|}
\hline
Parameters
&
CMB+SBL
&
{
\renewcommand{\arraystretch}{1}
\begin{tabular}{c}
CMB+SBL
\\
+LSS+$H_0$
\\
+CFHTLenS+PSZ
\end{tabular}
}
\\

\hline

$\Omega_{\rm b} h^2$ 	
			& $0.02273^{+0.00042}_{-0.00042}\,^{+0.00086}_{-0.00079}$ & $0.02267^{+0.00027}_{-0.00027}\,^{+0.00056}_{-0.00052}$ \\

$\Omega_{\rm cdm} h^2$ 
			& $0.131^{+0.006}_{-0.007}\,^{+0.014}_{-0.012}$           & $0.122^{+0.005}_{-0.007}\,^{+0.013}_{-0.010}$           \\

$\theta_{\rm s}$ 
			& $1.0397^{+0.0008}_{-0.0008}\,^{+0.0017}_{-0.0017}$      & $1.0406^{+0.0010}_{-0.0010}\,^{+0.0017}_{-0.0019}$      \\

$\tau$ 
			& $0.100^{+0.015}_{-0.016}\,^{+0.032}_{-0.029}$           & $0.083^{+0.013}_{-0.014}\,^{+0.029}_{-0.027}$           \\

$n_{\rm s}$ 
			& $0.996^{+0.016}_{-0.018}\,^{+0.034}_{-0.033}$           & $0.991^{+0.010}_{-0.011}\,^{+0.021}_{-0.019}$           \\

$\log(10^{10} A_s)$ 
			& $3.145^{+0.044}_{-0.043}\,^{+0.081}_{-0.086}$           & $3.107^{+0.032}_{-0.030}\,^{+0.064}_{-0.060}$           \\

$r$ 
			& $0.182^{+0.043}_{-0.050}\,^{+0.099}_{-0.085}$           & $0.206^{+0.042}_{-0.048}\,^{+0.097}_{-0.086}$           \\
\hline
$\ns$ 
			& $1.03^{+0.47}_{-0.50}\,^{+0.97}_{-0.95}$                & $0.66^{+0.33}_{-0.43};\,<1.35$                          \\

$m_s [\rm{eV}]$
			& $1.26^{+0.11}_{-0.16}\,^{+0.17}_{-0.27}$                & $1.26^{+0.11}_{-0.16}\,^{+0.17}_{-0.28}$                \\

$\tau_s[T_0]$
			& $<0.13 (1\sigma)$                                       & $>0.22 (1\sigma)$                                       \\

\hline
$\Delta\chi^2 (\text{A})$	&$-1.8$		&$-2.1$		\\
$\Delta\chi^2 (\text{B})$	&$-8.2$		&$-7.8$		\\
\hline
\end{tabular}
\end{center}
\caption{\label{tab:all}
Marginalized $1\sigma$ and $2\sigma$ confidence level limits for the cosmological parameters
obtained with the invisible sterile neutrino decays.
The decay lifetime $\tau_s$ is given in units of the age of the Universe $T_0$.
$\Delta\chi^2(\text{A})$ and $\Delta\chi^2(\text{B})$
give the variation of the cosmological
$\chi^2$ with respect to the fit of cosmological data without (A) and with (B) the SBL prior
for the mass of a stable sterile neutrino
(corresponding respectively to the grey and red regions and lines in the figures).
}
\end{table}

\begin{figure}
\centering
\includegraphics[width=\linewidth,page=5]{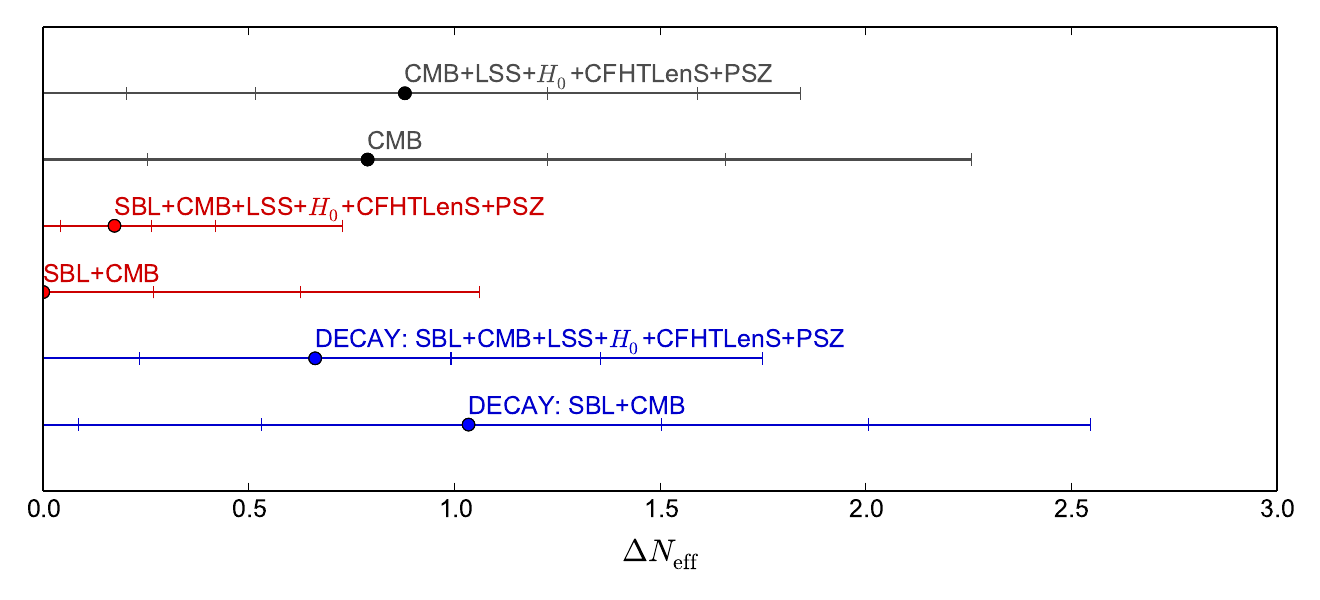}
\includegraphics[width=\linewidth,page=6]{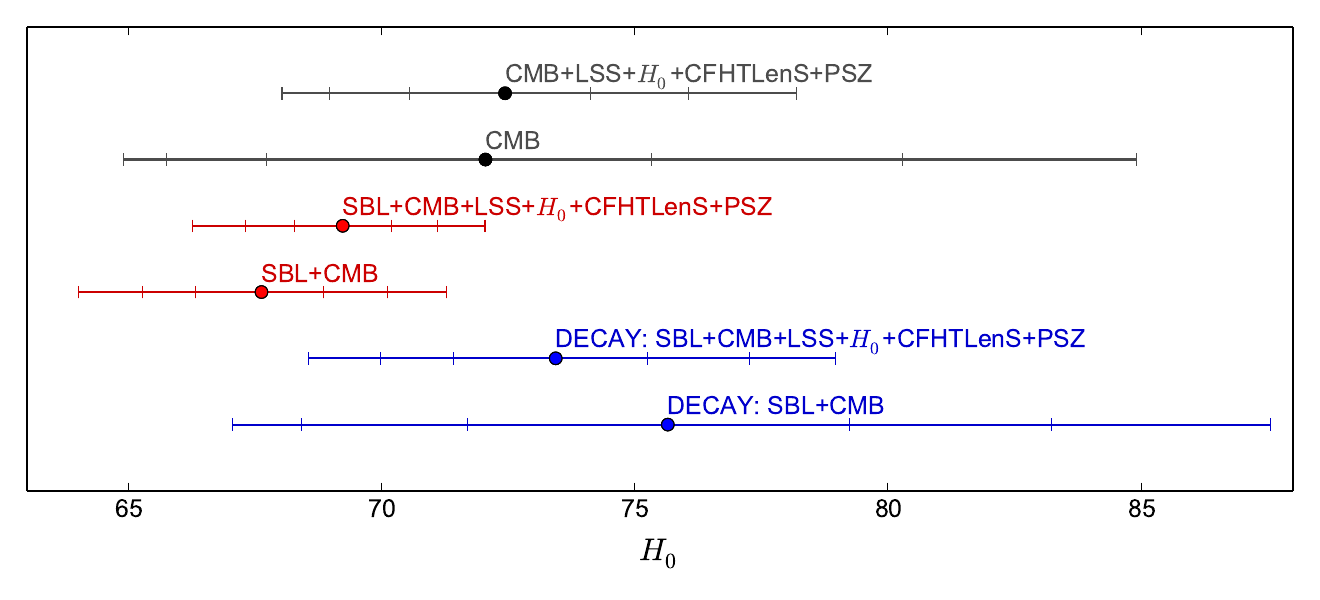}
\caption{\label{fig:int}
$1\sigma$, $2\sigma$ and $3\sigma$ marginalized error bars for $\ns$ and $H_0$
obtained in the different fits of the cosmological data considered in
Figs.~\ref{fig:cmb} and \ref{fig:all}.
The circles indicate the marginalized best fit value.
The black and red intervals are taken from Ref.~\cite{Archidiacono:2014apa}.
The blue intervals are obtained by adding the invisible sterile neutrino decays.
}
\end{figure}

In Fig.~\ref{fig:cmb}
we compared the allowed regions obtained
with the invisible decay of the sterile neutrino
with the corresponding regions
obtained in Ref.~\cite{Archidiacono:2014apa} for a stable sterile neutrino without and with the SBL prior.
One can see that the invisible decay of the sterile neutrino allows
$\Delta{N}_{\text{eff}} = 1$,
which corresponds to the full initial thermalization of the sterile neutrino,
even if the SBL prior forces the sterile neutrino mass to assume values around 1.2 eV.
In practice,
the invisible decay of the sterile neutrino
allows us to relax the upper bound of about 0.6 for
$\Delta{N}_{\text{eff}}$
obtained in Ref.~\cite{Archidiacono:2014apa}
with the SBL prior
and bring the allowed range of $\Delta{N}_{\text{eff}}$
at a level which is similar to that obtained in Ref.~\cite{Archidiacono:2014apa}
without the SBL prior
(see also \cite{Giusarma:2014zza,Zhang:2014dxk,Dvorkin:2014lea,Zhang:2014nta}).
This can also be seen
in the upper panel of Fig.~\ref{fig:int},
which shows the marginalized allowed interval of $\Delta{N}_{\text{eff}}$.

Figure~\ref{fig:cmb} shows also that by allowing the sterile neutrino to decay
one can recover a correlation between
$\Delta{N}_{\text{eff}}$
and
$H_0$
which is similar to that obtained in the analysis of CMB data without the SBL prior.
Hence,
we obtain that large values of $\Delta{N}_{\text{eff}}$
are correlated to large values of the Hubble constant $H_0$,
which are in agreement with the local measurements of $H_0$
(see \cite{Ade:2013zuv,Gariazzo:2013gua}).

From Tab.~\ref{tab:all}
one can see that small values of $\tau_s$ are preferred,
but the uncertainty is very large, because there is no bound at $2\sigma$.
The preference for small values of $\tau_s$ reflects the bound on a stable sterile neutrino mass
given by CMB data
(see the grey region in Fig.~\ref{fig:cmb}):
the bound can be evaded allowing $m_s \sim 1 \, \text{eV}$
if the sterile neutrino decays quickly.
However,
the uncertainty for $\tau_s$ is very large,
because of the wide range of allowed values of $\Delta{N}_{\text{eff}}$.
Hence, we performed also a run of \texttt{CosmoMC}
with $\Delta{N}_{\text{eff}}=1$,
which gave the more stringent upper bounds
$\tau_s < 0.11 \, T_0$ ($1\sigma$), $0.26 \, T_0$ ($2\sigma$),
where $T_0$ is the age of the Universe.

\begin{figure*}
\centering
\includegraphics[width=0.49\textwidth,page=3]{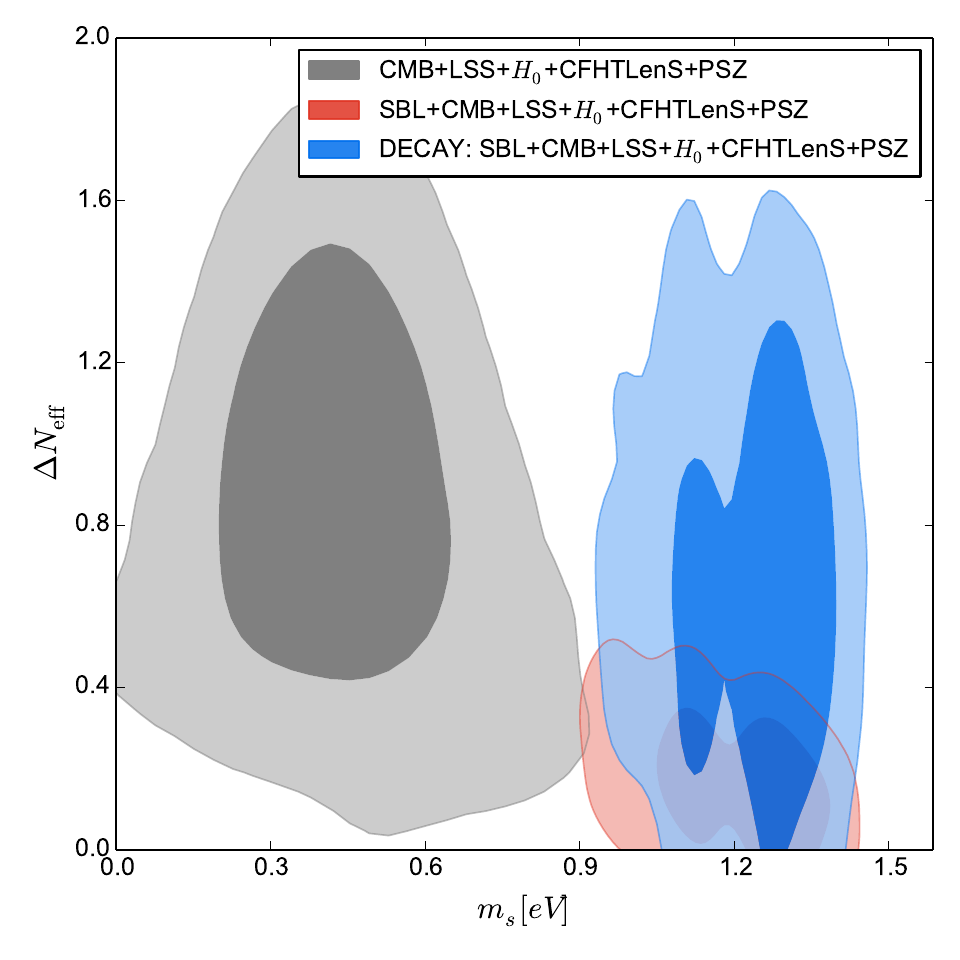}
\includegraphics[width=0.49\textwidth,page=4]{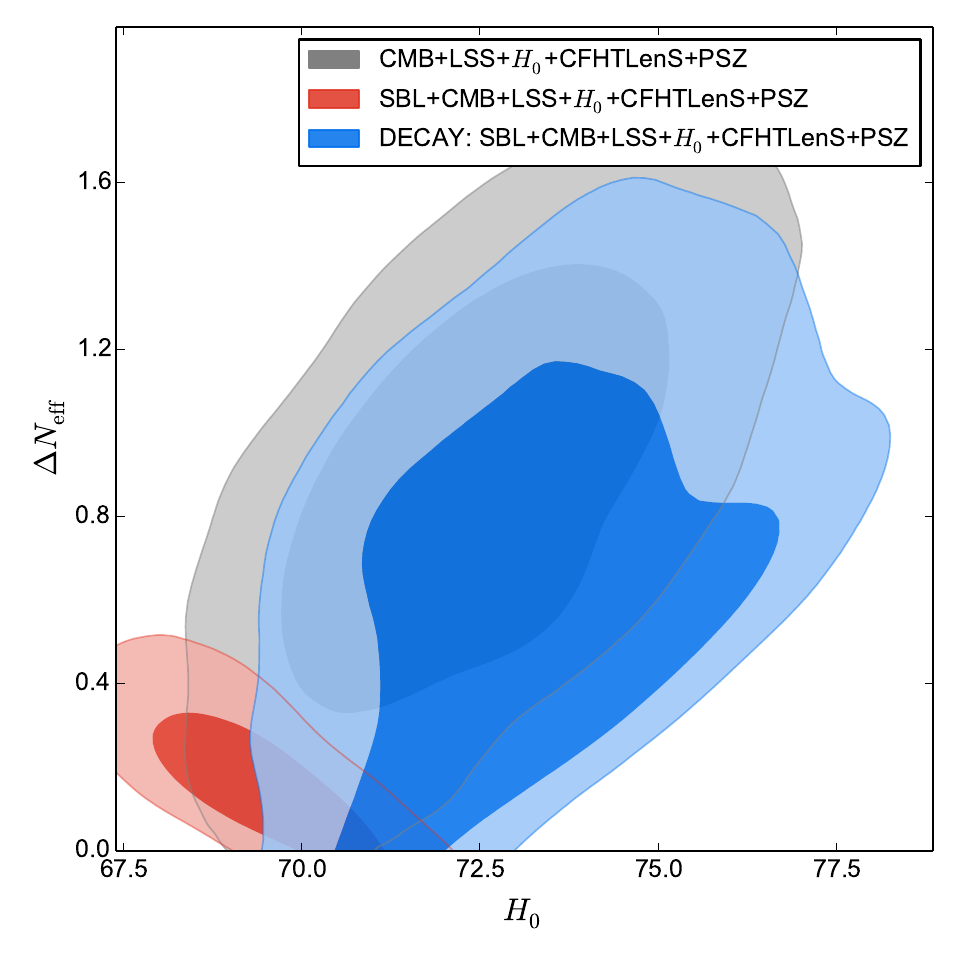}
\caption{\label{fig:all}
$1\sigma$ and $2\sigma$ marginalized allowed regions
obtained with the most complete cosmological data set considered in Ref.~\cite{Archidiacono:2014apa}
(the CMB data considered in Fig.~\ref{fig:cmb} plus LSS+$H_0$+CFHTLenS+PSZ).
The gray and red regions are those obtained in Ref.~\cite{Archidiacono:2014apa} without and with the SBL prior.
The blue regions are obtained by adding the invisible sterile neutrino decays.
}
\end{figure*}

Figure~\ref{fig:all}
shows the $1\sigma$ and $2\sigma$ marginalized allowed regions
corresponding to those of Fig.~\ref{fig:cmb}
and obtained by adding the same cosmological data considered in
Ref.~\cite{Archidiacono:2014apa}
on
Large Scale Structures (LSS),
local $H_0$ measurements,
cosmic shear (CFHTLenS)
and
the Sunayev-Zeldovich effect cluster counts from Planck (PSZ).
One can see that also this wide data set allows
$\Delta{N}_{\text{eff}}=1$
and the allowed range of
$\Delta{N}_{\text{eff}}$
is similar to that
obtained without the SBL prior
(see also Fig.~\ref{fig:int}).
As in Fig.~\ref{fig:cmb},
$\Delta{N}_{\text{eff}}$
and
$H_0$
are approximately correlated,
indicating relatively large values of $H_0$
for $\Delta{N}_{\text{eff}}=1$,
which are in agreement with the local measurements of $H_0$.

The $\Delta\chi^2$ values in Tab.~\ref{tab:all}
show that the fit of the cosmological data is excellent.
From Tab.~\ref{tab:all}
one can also see that the sterile neutrino lifetime $\tau_s$ has a $1\sigma$ lower bound,
without a $2\sigma$ bound.
In this case, large values of $\tau_s$ are preferred because
the free-streaming of a massive sterile neutrino can explain the
observed suppression of small-scale matter density fluctuations
\cite{Wyman:2013lza,Battye:2013xqa,Gariazzo:2013gua,Zhang:2014dxk,Dvorkin:2014lea,Archidiacono:2014apa,Zhang:2014nta}.
The large uncertainty on $\tau_s$ is again due to the
the wide range of allowed values of $\Delta{N}_{\text{eff}}$.
Assuming $\Delta{N}_{\text{eff}}=1$,
we obtained the more stringent bounds
$0.5 < \tau_s < 2.5 \, T_0$ ($1\sigma$)
and
$\tau_s > 0.14 \, T_0$ ($2\sigma$).

In conclusion,
we have proposed to solve the tension between
the fit of cosmological data with a sterile neutrino with the mass of about 1 eV indicated by
short-baseline neutrino oscillation data
and
the thermalization of this sterile neutrino in the early Universe
by introducing a decay of the sterile neutrino into invisible very light particles.
We have shown that a fit of the cosmological data with the SBL prior
allows a full thermalization of the sterile neutrino in the early Universe,
reconciling the sterile neutrino explanations of oscillations and cosmological data.
The decaying massive sterile neutrino has a beneficial effect for
the explanation of the observed suppression of small-scale matter density fluctuations
and leads to a larger value of the Hubble constant which is in agreement with
local measurements.

\begin{acknowledgments}
We would like to thank
M. Archidiacono,
N. Fornengo,
S. Hannestad,
A. Melchiorri,
Y.F Li
and
H.W. Long
for stimulating discussions
and fruitful collaboration in previous works.
\end{acknowledgments}


\end{document}